\documentclass[apjl]{emulateapj}
\usepackage[dvips]{color}

\def\ltsima{$\;\buildrel < \over \sim \;$}
\def\simlt{\lower.5ex \hbox{\ltsima}}
\def\gtsima{$\;\buildrel > \over \sim \;$}
\def\simgt{\lower.5ex \hbox{\gtsima}}

\shorttitle{Herschel/HIFI observations of IRC+10216}
\shortauthors{Neufeld et al.}

\begin{document}

\title{Herschel$^*$/HIFI observations of IRC+10216: \\
water vapor in the inner envelope of a carbon-rich AGB star }
\author{David A.~Neufeld\altaffilmark{1},
Eduardo~Gonz\'alez-Alfonso\altaffilmark{2}, Gary J. Melnick\altaffilmark{3},
Ryszard~Szczerba\altaffilmark{4}, Miroslaw~Schmidt\altaffilmark{4}, Leen Decin\altaffilmark{5}, Alex de Koter\altaffilmark{6,7}, Fredrik Sch\"oier\altaffilmark{8}, and Jos\'e Cernicharo\altaffilmark{9}}
\altaffiltext{*}{Herschel is an ESA space observatory with science instruments provided
by European-led Principal Investigator consortia and with important
participation from NASA}
\altaffiltext{1}{Department of Physics and Astronomy, Johns Hopkins University,
3400~North~Charles~Street, Baltimore, MD 21218, USA}
\altaffiltext{2}{Departamento de F\'{\i}sica, Universidad de Alcal\'a de Henares, Campus Universitario, E-28871 Alcal\'a de Henares, Madrid, Spain}
\altaffiltext{3}{Harvard-Smithsonian CfA, Cambridge, MA 02138, USA}
\altaffiltext{4}{N. Copernicus Astronomical Center, Toru{\'n}, Poland} 
\altaffiltext{5}{Instituut voor Sterrenkunde, K.~U.~Leuven, Belgium}
\altaffiltext{6}{Sterrenkundig Instituut Anton Pannekoek, University of Amsterdam, Science Park 904, NL-1098 Amsterdam, The Netherlands}
\altaffiltext{7}{Astronomical Institute, Utrecht University, The Netherlands}
\altaffiltext{8}{Onsala Space Observatory, Dept. of Radio and Space Science, Chalmers  University of Technology, SE--43992 Onsala, Sweden}
\altaffiltext{9}{CAB, INTA-CSIC, Ctra de Torrej\'on a Ajalvir, Madrid, Spain}

\begin{abstract}

We report the results of observations of ten rotational transitions of water vapor toward the carbon-rich AGB (asymptotic giant branch) star IRC+10216 (CW Leonis), carried out with {\it Herschel}'s HIFI instrument.  Each transition was securely detected by means of observations using the dual beam switch mode of HIFI.
The measured line ratios imply that water vapor is present in the inner outflow at small distances ($\le \rm few \, \times \, 10^{14}\, cm$) from the star, confirming recent results reported by Decin et al.\ from observations with Herschel's PACS and SPIRE instruments.  This finding definitively rules out the hypothesis that the observed water results from the vaporization of small icy objects in circular orbits.  The origin of water within the dense C-rich envelope of IRC+10216 remains poorly understood.   We derive upper limits on the H$_2^{17}$O/H$_2^{16}$O and H$_2^{18}$O/H$_2^{16}$O isotopic abundance ratios of $\sim 5 \times 10^{-3}$ (3 $\sigma$), providing additional constraints on models for the origin of the water vapor in IRC+10216.

\end{abstract}

\keywords{circumstellar matter -- stars: AGB and post-AGB -- stars: abundances }

\section{Introduction}

The carbon-rich AGB (asymptotic giant branch) star IRC+10216 (CW Leonis) is surrounded by a dense, outflowing envelope that harbors a rich chemistry.  Because of its proximity to Earth ($D \sim 100 - 170$~pc) and its large mass-loss rate ($\sim 1.5 - 3 \times 10^{-5} \rm M_\odot \, yr^{-1}$; see Crosas \& Menten 1997 and Sch{\"o}ier 
\& Olofsson 2000), IRC+10216 has proven to be the most valuable source available for the study of chemistry in a carbon-rich astrophysical environment; more than 60 molecules (e.g. He et al.\ 2008) have been detected within its circumstellar envelope (CSE), made up of the elements H, C, N, O, F, Na, Mg, Al, Si, S, P, F, and K.  The photosphere of IRC+10216 has been significantly enriched in carbon -- owing to the dredge-up of helium-burning nuclear reaction products from the core -- leading to a large C/O ratio ($\sim 1.4$; Winters, Dominik, \& Sedlmayr 1994) and giving rise to a circumstellar chemistry that is qualitatively different from that typical of oxygen-rich AGB stars. In particular, while the most abundant molecules in O-rich CSEs are CO and H$_2$O, those in IRC+10216 are CO, HCN, and $\rm C_2H_2$.  Prior to 2001, the only oxygen-rich molecules reported in the CSE of IRC+10216 were CO, SiO, and HCO$^+$ (Lucas \& Gu{\'e}lin 1999), but in the last decade four other O-bearing molecules have been detected: H$_2$O (Melnick et al.\ 2001), OH (Ford et al.\ 2003), $\rm H_2CO$ (Ford et al.\ 2004), and $\rm C_3O$ (Tenenbaum et al.\ 2006).  

The discovery of water vapor in IRC+10216, obtained by observations of the $1_{10}-1_{01}$ rotational transition near 557 GHz with the use of the {\it Submillimeter Wave Astronomy Satellite} (SWAS), was puzzling because models for thermochemical equilibrium in the photospheres of carbon-rich stars predict water abundances of only $\sim 10^{-10}$ relative to H$_2$ (e.g.\ Cherchneff 2006), many orders of magnitude below the detection threshold or the abundance $\sim 10^{-7}$ inferred (Ag{\'u}ndez and Cernicharo 2006; Gonz{\'a}lez-Alfonso, Neufeld and Melnick 2007) from SWAS observations.  Several hypotheses have been proposed for the origin of the observed water vapor.  Ford \& Neufeld (2001) proposed a model whereby the water vapor was released into the outflow by the vaporization of a Kuiper belt analog, in which orbiting icy objects -- heated by the enhanced luminosity of the AGB star -- underwent sublimation.  Willacy (2004) subsequently suggested that the observed water could be formed from CO and H$_2$ by means of Fischer-Tropsch catalysis on metallic dust grains, whilst Ag{\'u}ndez and Cernicharo (2006; hereafter AC06) proposed that water might be formed in the outer envelope by means of radiative association of H$_2$ with O atoms produced by the photodissociation of $\rm ^{13}CO$ exposed to the ultraviolet interstellar radiation field (ISRF).  Because SWAS had access to only a single transition of water vapor, it was unable to place observational constraints that could readily distinguish between these hypotheses.

In a theoretical study (Gonz{\'a}lez-Alfonso, Neufeld and Melnick 2007; hereafter GNM) performed in anticipation of the {\it Herschel Space Observatory}, it was discussed how each of these hypotheses gave rise to a specific prediction for the spatial distribution of the emitting water vapor.  GNM investigated how multitransition observations with {\it Herschel} might be used to determine that distribution, and showed, in particular, that the relative strength of lines of higher excitation than those accessible to SWAS was a decreasing function of the inner radius, $R_{\rm in}$, of the region containing water vapor.  The vaporization of a Kuiper belt analog would lead to 
$R_{\rm in} \sim 2 \times 10^{15}{\rm \, cm} \sim 30 \,R_*$ (Model B in GNM; where $R_* \sim 8 \times 10^{13}\, \rm cm$ is the assumed stellar radius)  because any icy object (at least of small size and on a circular orbit) within that radius would have been vaporized before IRC+10216 ascended the AGB.  Water production by means of Fischer-Tropsch catalysis would result in a similar value of $R_{\rm in}$, according to calculations of Willacy (2004).  The formation of water following the photodissociation of CO in an outer shell, as suggested by AC06, by contrast, would lead to a much larger $R_{\rm in} \sim 4.3 \times 10^{16} {\rm \, cm} \sim 500 \, R_*$ (Model C in GNM), and considerably smaller strengths for the higher-excitation water transitions.  Conversely, were $R_{\rm in}$ significantly smaller than $2 \times 10^{15}\rm \, cm$ -- a possibility that had not been anticipated by any {\it specific} model -- then the high-excitation transitions would be relatively stronger (Model A in GNM).

In this {\it Letter}, we report the results obtained from {\it Herschel} (Pilbratt et al.\ 2010) observations of 10 rotational transitions of water vapor, carried with the Heterodyne Instrument for the Far Infrared (HIFI; de Graauw et al.\ 2010).  The observations and data reduction are described in \S 2, and the spectral line profiles and line intensities are presented in \S 3.  In \S 4, we discuss the spatial distribution inferred for the water vapor in the CSE of IRC+10216, in the context of various hypotheses for its origin.  We compare our results with those reported recently in an entirely independent study performed by Decin et al. (2010; hereafter D10) with the use of the PACS and SPIRE instruments on {\it Herschel}.

\section{Observations and data reduction}

Observations of IRC+10216 were carried out in May 2010 as part of the HIFISTARS Key Program.  We used the HIFI instrument in dual beam switch (DBS) mode to target 10 rotational transitions of water vapor.  The list of observed transitions, and the details of each observation, are given in Table 1.    The telescope beam was centered on IRC+10216 at coordinates $\alpha=\rm 9h\, 47m \, 57.38s$, $\delta= +13^0 16^\prime 43.7^{\prime \prime}$ (J2000), and the reference positions were located at offsets of 3$^\prime$ on either side of the source.  The data were processed using the standard HIFI pipeline to Level 2, providing fully calibrated spectra with the intensities expressed as antenna temperature and the frequencies in the frame of the Local Standard of Rest (LSR).
For five of the ten observed transitions, we used a pair of local oscillator (LO) frequencies, separated by a small offset, to confirm the assignment of any observed spectral feature to either the upper or lower sideband of the (double side band) HIFI receivers.
The resultant spectra were coadded so as to recover the signal-to-noise ratio that would have been obtained at a single LO setting.  Spectra obtained for the horizontal and vertical polarizations were found to be very similar in their appearance and noise characteristics and were likewise coadded.
\vskip 0.2 true in

\begin{figure}
\includegraphics[scale=0.48]{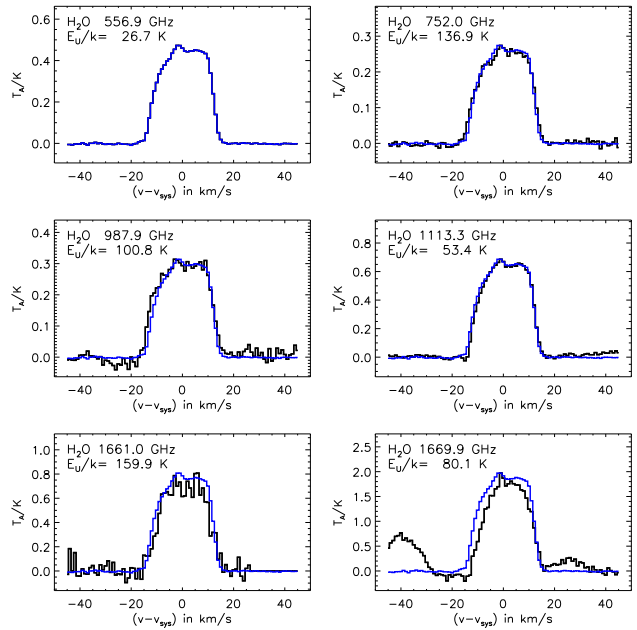}
\caption{Herschel/HIFI spectra of six rotational transitions of H$_2$O with upper state energies $E_{\rm U}/k \le 160$~K.  The blue line superposed on each spectrum shows the profile of the $1_{10}-1_{01}$ ground-state transition of ortho-water (top left panel).  Doppler velocities are expressed relative to the LSR velocity of the source, taken as $-25.5 \, \rm km \,s^{-1}$.}
\end{figure}

\begin{figure}
\includegraphics[scale=0.48]{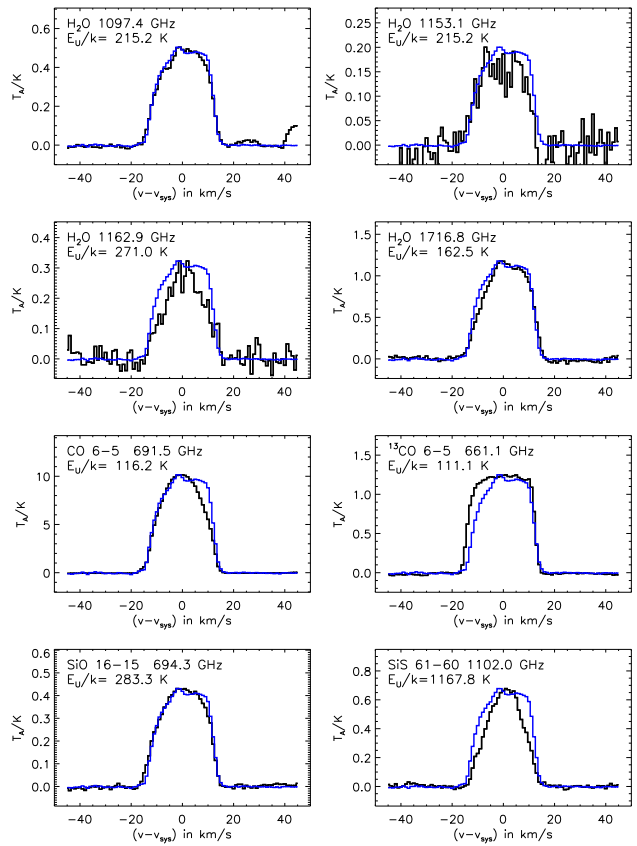}
\caption{same as Fig.\ 1, except for four rotational transitions of H$_2$O with upper state energies $E_{\rm U}/k \ge 160$~K, and one rotational transition each of CO, $^{13}$CO, SiO and SiS.}
\end{figure}

\section{Results}

The integrated antenna temperatures and line fluxes are given for each detected transition of water vapor in Table 2, together with the energy of the upper state.
The HIFI instrument used for these observations provides a spectral resolution (1.1~MHz for the Wide Band Spectrometer) that is much superior to that of the PACS or SPIRE instruments; this property of HIFI typically provides for the unambiguous identification of strong lines, and yields spectral line profiles that can be used to probe the kinematics of the outflowing gas.   In Figures 1 and 2, we present the
spectra obtained for each of the ten detected water vapor transitions, each smoothed to a spectral resolution of $\rm 1 \, km \, s^{-1}$ to improve the signal-to-noise ratio.   The lower four panels in Figure 2 show comparison spectra for one transition each of the molecules CO, $^{13}\rm CO$, SiO and SiS; taking advantage of the wide bandwidth of the HIFI receivers (4 GHz for $\nu \le 1250$~GHz and 2.4 GHz for $\nu \ge 1410$~GHz), we observed these lines simultaneously with nearby water transitions.  In Figures 1 and 2, the blue line superposed on each spectrum shows the profile of the $1_{10}-1_{01}$ ground-state transition of ortho-water.  

Although a full analysis of the spectral line profiles will be the subject of a future publication, several features are obvious from an inspection of the spectra.  Most of the water lines share a distinctive profile that is quite dissimilar from those of either the $^{12}$CO or $^{13}$CO transitions.  In particular, the blue sides of the water lines are very similar to the nearly parabolic profile shown by $\rm ^{12}CO$, while the red sides are more similar to that of the $\rm ^{13}CO$ line, which shows the more rectangular profile expected of an optically-thin line.  In these respects, the typical water line profiles are most similar to that of the silicon monoxide line shown in Figure 2.  The observed asymmetry in the line profiles likely reflects an asymmetry in the distribution of the emitting H$_2$O and SiO, perhaps suggesting a common origin\footnote{Cherchneff (2006) has proposed pulsation-driven shock waves as the source of SiO in carbon-rich AGB stars, and -- although the models presented in that 2006 study failed to predict any enduring enhancement in the water abundance behind such shock waves -- preliminary results from a more recent calculation do indeed suggest (Cherchneff 2010) that the observed water vapor could be produced along with SiO by shocks in the inner envelope.}.  The highest excitation water line to have been detected in our study, the 1162.912~GHz transition, exhibits a profile that is rather different from the typical water line profile, with a narrow core that is similar to that present in the SiS 1102.029~GHz line profile; here, emission from the acceleration zone may be implicated.  In addition, the other water lines typically show a small narrow emission bump near the systemic velocity of the source, perhaps also representing emission from material that has not yet been fully accelerated.

\begin{deluxetable}{cccclr}

\tablewidth{0pt}
\tablecaption{Observations of IRC+10216} 
\tablehead{Transition & $\nu$ & Mixer & HPBW$^a$  & Date & $t_{\rm obs}^b$ \\
                       & (GHz)   & band  & ($^{\prime \prime}$)     & 2010 & (s)}

\startdata
$1_{10}-1_{01}$~(o)$^c$ & \phantom{1}556.936 & 1b & 38 & May 4, 11 & $3150^d$ \\
$2_{11}-2_{02}$~(p)\phantom{$^c$} & \phantom{1}752.033 & 2b & 28 & May 12 & $3236^d$ \\
$2_{02}-1_{11}$~(p)\phantom{$^c$} & \phantom{1}987.927 & 4a & 22 & May 16 & $3242^d$ \\
$3_{12}-3_{03}$~(o)\phantom{$^c$} & 1097.365 & 4b & 19 & May 11 & $10446^d$ \\
$1_{11}-0_{00}$~(p)\phantom{$^c$} & 1113.343 & 4b & 19 & May 11 & $3200^d$ \\
$3_{12}-2_{21}$~(o)\phantom{$^c$} & 1153.127 & 5a & 18 & May 12 & $1539^e$ \\
$3_{21}-3_{12}$~(o)\phantom{$^c$} & 1162.912 & 5a & 18 & May 12 & $1539^e$ \\
$2_{21}-2_{12}$~(o)\phantom{$^c$} & 1661.008 & 6b & 13 & May 14 & $1615^f$ \\
$2_{12}-1_{01}$~(o)\phantom{$^c$} & 1669.905 & 6b & 13 & May 14 & $1615^f$ \\
$3_{03}-2_{12}$~(o)\phantom{$^c$} & 1716.770 & 7a & 12 & May 14,15 & $10412^d$ 
\enddata

\tablenotetext{a}{Half power beam width}
\tablenotetext{b}{Total observing time, including overheads }
\tablenotetext{c}{The letters o and p indicate whether the transition is of ortho- or of para-water}\tablenotetext{d}{Data obtained with two separate LO settings of duration $t_{\rm obs}/2$ }
\tablenotetext{e,f}{Pairs of lines observed simultaneously } 
\end{deluxetable}

\section{Discussion}

In Figure 3, we compare the measured line fluxes tabulated in Table 2 with the predictions of the GNM models.  Here, the observed line fluxes are represented by black crosses for each of the ten detected transitions, ordered from left to right in increasing energy of the upper state, and the GNM predictions are shown by the solid lines for various values of the inner radius, $R_{\rm in}$, of the water emitting region.
These predictions were ``calibrated" by adjusting the assumed water abundance to match the 557 GHz line flux detected by SWAS.  Clearly, the best fit to the data is obtained for the smallest value of $R_{\rm in}$ considered by GNM, $4.5 \times 10^{14}\, \rm cm$ ($\sim 5.6 \, R_*$), but even for that value the model {\it underpredicts} the relative strengths of the transitions of highest excitation.  Thus, our HIFI observations confirm the inference drawn by D10 -- from the detection of high-lying water rotational lines with PACS and SPIRE -- about the spatial distribution of water vapor; water is clearly present at distances smaller than $4.5 \times 10^{14}\, \rm cm$ ($\sim 5.6 \, R_*$) from the star.  The presence of water that close to the star definitively falsifies the model proposed by Ford \& Neufeld (2001), in which the origin of the observed water vapor was the vaporization of icy comets in a Kuiper belt analog, because any icy object (at least of small size and on a circular orbit) within $4.5 \times 10^{14}\, \rm cm$  would have been vaporized before IRC+10216 ascended the AGB.  The models proposed by Willacy (2004; i.e. Fischer-Tropsch catalysis) and particularly AC06 (production via radiative association in an outer shell) are similarly excluded.  The fact that the 1113~GHz / 557~GHz line ratio is in good agreement with the GNM model implies that the ortho-to-para ratio is close to 3, the value assumed in the model.  The dashed black line in Figure 3 shows our best-fit model, with $R_{\rm in}= 1.0 \times 10^{14}\, \rm cm$ ($\sim 1.3 \, R_*$), and a slightly larger water abundance $\rm H_2O/H_2 = 8.1 \times 10^{-8}$ than that assumed in GNM, now chosen to optimize the fit to {\it all} the transitions observed with HIFI.

\begin{figure}
\includegraphics[scale=0.50]{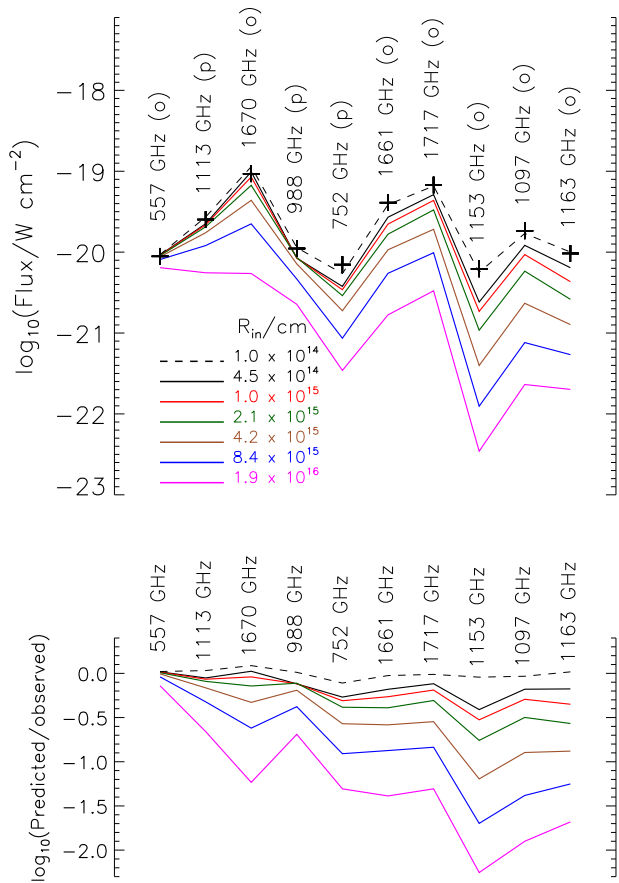}

\caption{Comparison of the measured water line fluxes (black crosses) with predictions of the GNM models for various values of the inner radius, $R_{\rm in}$, of the water-emitting region. The ten rotational transitions appear from left to right in order of increasing energy for the upper state (except for the 1153 GHz and 1097 GHz transitions which originate in the same upper state).}
\end{figure}

\begin{deluxetable}{ccccc}\tablewidth{0pt}
\tablecaption{Water line fluxes measured toward IRC+10216}
\tablehead{Transition & $\nu$ & $E_U$/k & $\int T_A dv$  & Flux$^{a,b}$ \\
                       & (GHz)   & (K) & ($\rm K\,km\,s^{-1}$) & ($\rm 10^{-20}\,W \, cm^{-2}$)} 
\startdata
$1_{10}-1_{01}$~(o)$^c$ & \phantom{1}556.936 & \phantom{1}26.7& 10.3 & 0.89\\
$2_{11}-2_{02}$~(p)\phantom{$^c$} & \phantom{1}752.033 & 136.9& \phantom{1}6.0 & 0.70\\
$2_{02}-1_{11}$~(p)\phantom{$^c$} & \phantom{1}987.927 & 100.8& \phantom{1}7.2 & 1.11\\
$3_{12}-3_{03}$~(o)\phantom{$^c$} & 1097.365 & 215.2 & 10.7 & 1.83\\
$1_{11}-0_{00}$~(p)\phantom{$^c$} & 1113.343 & \phantom{1}53.4 & 14.6 & 2.53 \\
$3_{12}-2_{21}$~(o)\phantom{$^c$} & 1153.127 & 215.2 & \phantom{1}3.4 & 0.62 \\
$3_{21}-3_{12}$~(o)\phantom{$^c$} & 1162.912 & 271.0 & \phantom{1}5.3 & 0.96 \\
$2_{21}-2_{12}$~(o)\phantom{$^c$} & 1661.008 & 159.9 & 15.5 & 4.09\\
$2_{12}-1_{01}$~(o)\phantom{$^c$} & 1669.905 & \phantom{1}80.1 & 35.1 & 9.56\\
$3_{03}-2_{12}$~(o)\phantom{$^c$} & 1716.770 & 162.5 & 24.2 & 6.74
\enddata
\tablenotetext{a}{For an unresolved source at the beam center}
\tablenotetext{b}{We conservatively estimate the flux uncertainty to be less than 15$\%$.}
\tablenotetext{c}{The letters o and p indicate whether the transition is of ortho- or of para-water}
\end{deluxetable}

At present, the origin of water vapor in IRC+10216 remains poorly-understood.  D10 have proposed an alternative model -- discussed in greater detail by Ag{\'u}ndez, Cernicharo and Gu{\'e}lin (2010; hereafter ACG) --  involving photochemistry in the inner envelope.  As in the AC06 model, oxygen atoms are generated by photodissociation of $\rm ^{13}CO$ and SiO by the ultraviolet ISRF, but -- unlike the AC06 model -- the UV radiation is assumed to penetrate deeply into a clumpy CSE.  In the D10/ACG model, the oxygen atoms are liberated close to the star, where the temperature is sufficient to drive H$_2$O production via a sequence of two H-atom extraction reactions with activation energy barriers: $\rm O(H_2,H)OH(H_2,H)H_2O$.  Because the mean visual extinction through the CSE is $\sim 100$~mag, this scenario requires the existence of channels of greatly reduced extinction through which the ultraviolet radiation can penetrate.  The consequences of this model for the abundances and spatial distribution of the many other molecules detected in IRC+10216, many of which have been observed interferometrically, has yet to be fully investigated.   

One possible test of this model might involve a search for H$_2^{17}$O or H$_2^{18}$O.  Because the $\rm ^{12}CO$ photodissociation rate is sharply reduced by self-shielding, D10 have emphasized the importance of $\rm ^{13}CO$ as a source of atomic O that can react to form H$_2$O.  The photodissociation rates for  C$^{17}$O and C$^{18}$O would presumably be at least as large as that for $\rm ^{13}CO$.  Thus, if $\rm ^{13}CO$, C$^{17}$O and C$^{18}$O were the only suppliers of atomic oxygen, the $\rm H_2^{16}O/H_2^{18}O$ and $\rm H_2^{16}O/H_2^{17}O$ ratios would approach the $\rm ^{13}C$O/C$^{18}$O and $\rm ^{13}C$O/C$^{17}$O ratios respectively.  Given the $\rm ^{13}C/^{12}C$, $\rm ^{18}O/^{16}O$ and $\rm ^{17}O/^{16}O$ isotopic ratios determined by Kahane et al.\ (1992) and by Cernicharo, Gu{\'e}lin \& Kahane (2000) for IRC+10216, the $\rm ^{13}CO$/C$^{17}$O and $\rm ^{13}CO$/C$^{18}$O ratios are respectively $\sim$ 18 and 28, each a factor of $45$ (= $\rm ^{12}C/^{13}C$) smaller than the elemental $\rm ^{16}O$/$^{17}$O and $\rm ^{16}O$/$^{18}$O ratios in the CSE.   If SiO or $\rm ^{12}CO$ are significant additional sources of atomic oxygen in the D10/ACG picture, then a $\rm H_2^{16}O$/H$_2^{17}$O ratio larger than 18 -- or a $\rm H_2^{16}O$/H$_2^{18}$O ratio larger than 28 -- might still be consistent with the model.

Data obtained in a full HIFI spectral survey carried out toward IRC+10216 (Cernicharo et al.\ 2010) place upper limits on the flux of the H$_2^{17}$O (552.021 GHz) and H$_2^{18}$O (547.676 GHz) $1_{10}-1_{01}$ transitions.  Comparing these with the observed flux in the 556.936~GHz H$_2^{16}$O $1_{10}-1_{01}$ transition, we determined that the spectral survey places 3~$\sigma$ lower limits on both the 556.936~GHz/552.021~GHz and 556.936~GHz/547.676~GHz line ratios of $\sim 100$.  Taking account of optical depth effects, we find that these correspond to a lower limit of $\sim 200$ on both the $\rm H_2^{16}O/H_2^{17}O$ and $\rm H_2^{16}O/H_2^{18}O$ abundance ratios.  Given the elemental isotopic ratios $\rm ^{16}O/^{18}O \sim 1260$ and $\rm ^{16}O/^{17}O \sim 840$ (Kahane et al.\ 1992), these limits imply that the abundance of the minor isotopologues could only be enhanced (by means of isotope-selective photodissociation of CO, for example) by at most a factor 4 (H$_2^{17}$O) or 6 (H$_2^{18}$O).   Detailed modeling will be required to determine whether our non-detections of H$_2^{17}$O and H$_2^{18}$O are consistent with the water production mechanism proposed by D10/ACG.

\acknowledgments

HIFI has been designed and built by a consortium of institutes and university departments from across
Europe, Canada and the United States under the leadership of SRON Netherlands Institute for Space
Research, Groningen, The Netherlands and with major contributions from Germany, France and the US.
Consortium members are: Canada: CSA, U.~Waterloo; France: CESR, LAB, LERMA, IRAM; Germany:
KOSMA, MPIfR, MPS; Ireland, NUI Maynooth; Italy: ASI, IFSI-INAF, Osservatorio Astrofisico di Arcetri-
INAF; Netherlands: SRON, TUD; Poland: CAMK, CBK; Spain: Observatorio Astron\'omico Nacional (IGN),
Centro de Astrobiolog\'a (CSIC-INTA). Sweden: Chalmers University of Technology - MC2, RSS \& GARD;
Onsala Space Observatory; Swedish National Space Board, Stockholm University - Stockholm Observatory;
Switzerland: ETH Zurich, FHNW; USA: Caltech, JPL, NHSC.

This research was performed, in part, through a JPL contract funded by the National Aeronautics and Space Administration.
E.G-A  is a Research Associate at the Harvard-Smithsonian 
Center for Astrophysics.
R.~Sz. and M.~Sch.\ acknowledge support from grant N~203~581040.


\end{document}